\documentclass[preprint,prb,aps,floats,amssymb,showpacs,draft,%
floatfix,superscriptaddress]{revtex4} 
\usepackage{psfig}

\begin{document}

\title{ 
Multicritical behavior of two-dimensional 
anisotropic antiferromagnets in a magnetic field.  } 
\author{Andrea Pelissetto} 
\affiliation{Dipartimento di Fisica
  dell'Universit\`a di Roma ``La Sapienza" and INFN, Roma, Italy.}
\author{Ettore Vicari} 
\affiliation{ 
Dipartimento di Fisica dell'Universit\`a di Pisa and INFN, Pisa,
  Italy.  } 

\date{\today}

\begin{abstract}
We study the phase diagram and multicritical behavior of anisotropic
Heisenberg antiferromagnets on a square lattice in the presence of a
magnetic field along the easy axis. We argue that, beside the Ising
and XY critical lines, the phase diagram presents a first-order
spin-flop line starting from $T=0$, as in the three-dimensional case.
By using field-theory methods, we show that the multicritical point
where these transition lines meet cannot be O(3) symmetric and occurs
at finite temperature. We also predict how the critical temperature of
the transition lines varies with the magnetic field and the uniaxial
anisotropy in the limit of weak anisotropy.
\end{abstract}

\pacs{64.60.Kw, 05.10.Cc, 05.70.Jk, 75.10.Hk}

\maketitle


\section{Introduction}
\label{intro}

Anisotropic antiferromagnets in an external magnetic field have been
studied for a long time.  In order to determine their phase diagram,
they have often been modelled by using the classical XXZ model
\begin{equation}
{\cal H} = 
J \sum_{\langle mn \rangle} \vec{S}_n\cdot \vec{S}_m + 
A \sum_{\langle mn \rangle} S_{m,z} S_{n,z} - H \sum_m S_{m,z},
\label{xxz}
\end{equation}
where $\langle mn \rangle$ indicates a nearest-neighbor pair. 
Equivalently, one can use the Hamiltonian
\begin{equation}
{\cal H} = 
J \sum_{\langle mn \rangle} \vec{S}_n\cdot \vec{S}_m + 
D \sum_{m} S_{m,z}^2 - H \sum_m S_{m,z},
\end{equation}
with a single-ion anisotropy term.  The most interesting case
corresponds to uniaxial systems that show a complex phase
diagram. They correspond to Hamiltonians with $A > 0$ or $D <
0$. Several quasi-two-dimensional uniaxial antiferromagnets have been
studied experimentally, such as K$_2$MnF$_4$, Rb$_2$MnF$_4$,
Rb$_2$MnCl$_4$.
\cite{DRRH-82,DD-85,GJ-86,REPRKG-86,CABPST-93,KST-98,CLBE-01}

\begin{figure}[tb]
\centerline{\psfig{width=9truecm,angle=0,file=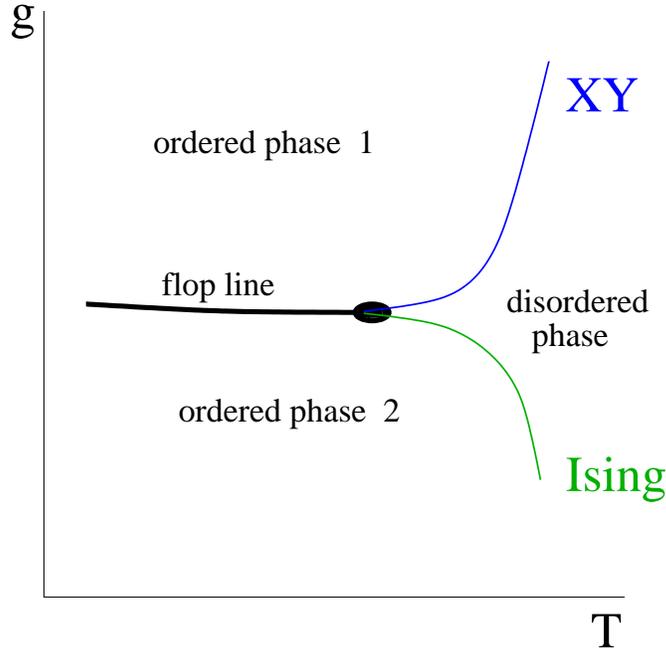}}
\vspace{2mm}
\caption{Phase diagram with a bicritical point where an Ising and an XY line meet.
Here, $T$ is the temperature and $g$  a second relevant parameter.
The thick black line (``flop line") corresponds to a first-order transition.
}
\label{bicri}
\end{figure}

\begin{figure}[tb]
\centerline{\psfig{width=9truecm,angle=0,file=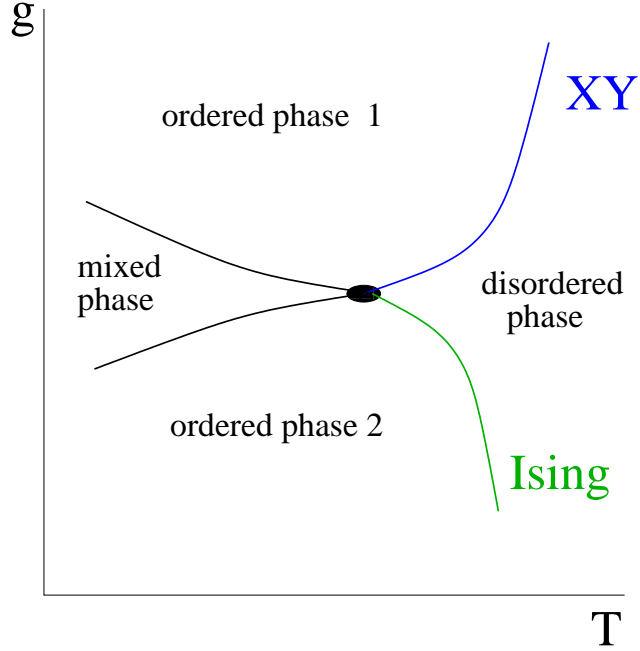}}
\vspace{2mm}
\caption{
Phase diagram with a tetracritical point where an Ising and an XY line
meet.  Here, $T$ is the temperature and $g$ a second relevant
parameter.  }
\label{tetra}
\end{figure}

\begin{figure}[tb]
\centerline{\psfig{width=9truecm,angle=0,file=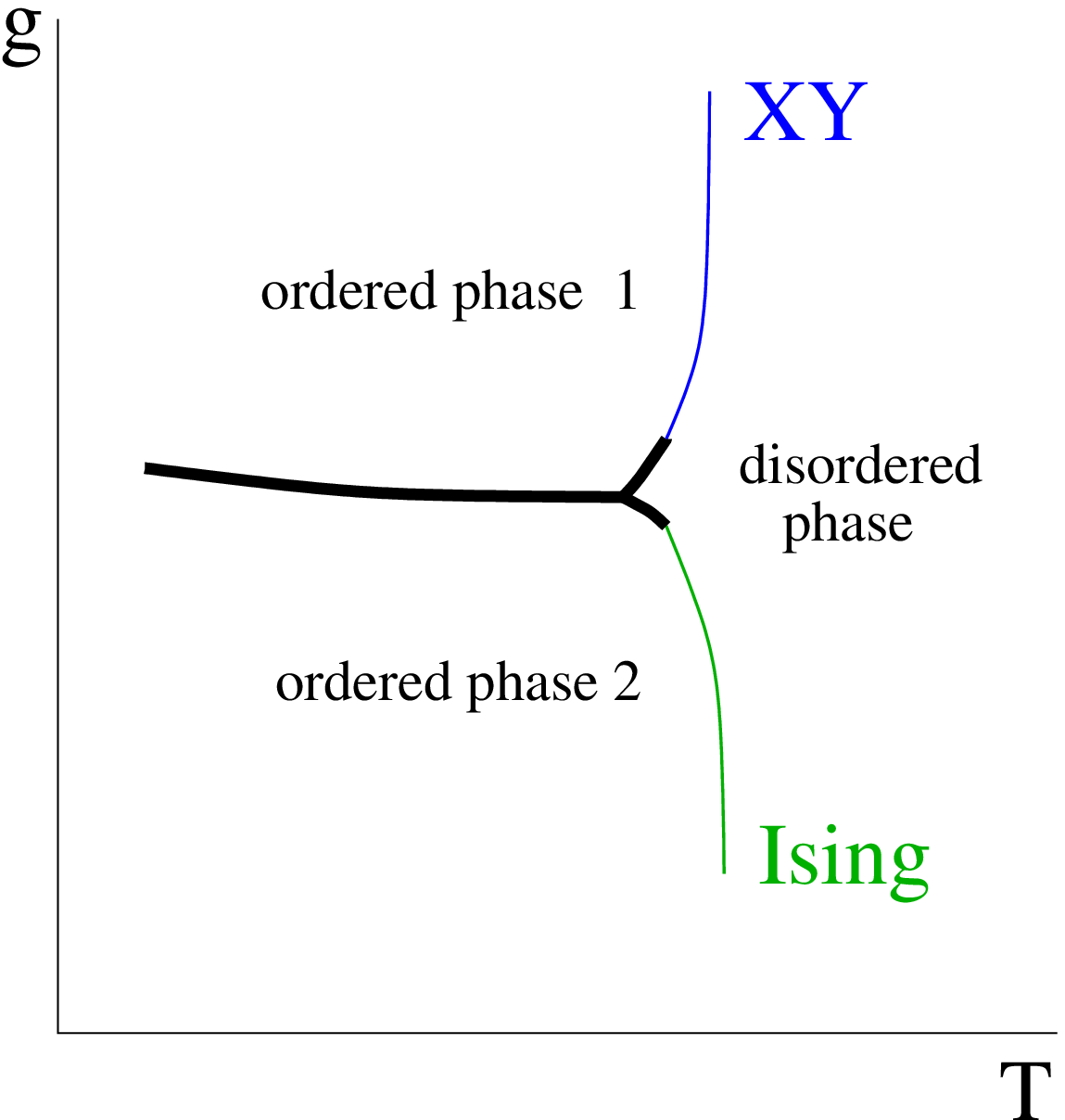}}
\vspace{2mm}
\caption{
Phase diagram with a bicritical point of first order.
The thick black lines correspond to first-order transitions.  }
\label{tricri}
\end{figure}

Some general features of the phase diagram of anisotropic
antiferromagnets are well known.  Their phase diagram in the $T$-$H$
plane presents two critical lines, belonging to the Ising and XY
universality class, respectively, which meet at a multicritical point
(MCP).  The nature of the MCP has been the object of several
theoretical studies. In three dimensions (3D), the issue has been recently
studied using a field-theory approach.\cite{CPV-03,HPV-05} The
starting point is the ${\rm O}(n_1)\oplus{\rm O}(n_2)$ symmetric
Landau-Ginzburg-Wilson (LGW) $\Phi^4$ theory \cite{KNF-76}
\begin{eqnarray}
{\cal H}_{\rm LGW} &=&\int d^d x \Bigl\{
{1\over 2}[(\partial_\mu \Phi_1)^2 +(\partial_\mu \Phi_2)^2] 
+ {1\over 2}[r_1 \Phi_1^{2}+ r_2 \Phi_2^{2}]
\nonumber \\ 
&& 
+ {1\over 4!} [ u_{10}(\Phi_1^{2})^2 + u_{20} (\Phi_2^{2})^2] + 
{1\over 4} w_0 \Phi_1^{2} \Phi_2^{2} \Bigr\},
\label{mcphi4}
\end{eqnarray}
where $\Phi_1$ and $\Phi_2$ are vector fields with $n_1$ and $n_2$ real
components, respectively. In our case $n_1=2$ and $n_2=1$ so that
Hamiltonian (\ref{mcphi4}) is symmetric under ${\mathbb Z}_2\oplus
{\rm O}(2)$ transformations.  The RG flow has been studied by
computing and analyzing high-order perturbative
expansions,\cite{CPV-03,HPV-05} to five and six loops.  It has been
shown that the stable fixed point (FP) of the theory is the biconal
FP, and that no enlargement of the symmetry to O(3) must be
asymptotically expected because the corresponding O(3) FP is unstable,
correcting earlier claims \cite{KNF-76} based on low-order
$\epsilon$-expansion calculations.  The perturbative results allow us
to predict that the transition at the MCP is either continuous and
belongs to the biconal universality class or that it is of first
order---this occurs if the system is not in the attraction domain of
the stable biconal FP. Which of these two possibilities occurs is
still an open issue.  A mean-field analysis \cite{LF-72,KNF-76} shows
that the MCP is bicritical if $\delta_0\equiv u_{10} u_{20} - 9 w_0^2
< 0$, and tetracritical if $\delta_0>0$.  Figs.~\ref{bicri} and
\ref{tetra} sketch the corresponding phase diagrams.  At one loop in
the $\epsilon$ expansion the biconal FP is associated with a
tetracritical phase diagram.\cite{KNF-76} Since uniaxial
antiferromagnets have a bicritical phase diagram, this result predicts
a first-order MCP as in Fig.~\ref{tricri}. First-order transitions are
also expected along the critical lines that separate the disordered
phase from the ordered ones. They start at the MCP, extend up to
tricritical points, and are followed by lines of XY and Ising
transitions.

\begin{figure*}[tb]
\begin{minipage}{7cm}
\centerline{\psfig{width=6truecm,angle=0,file=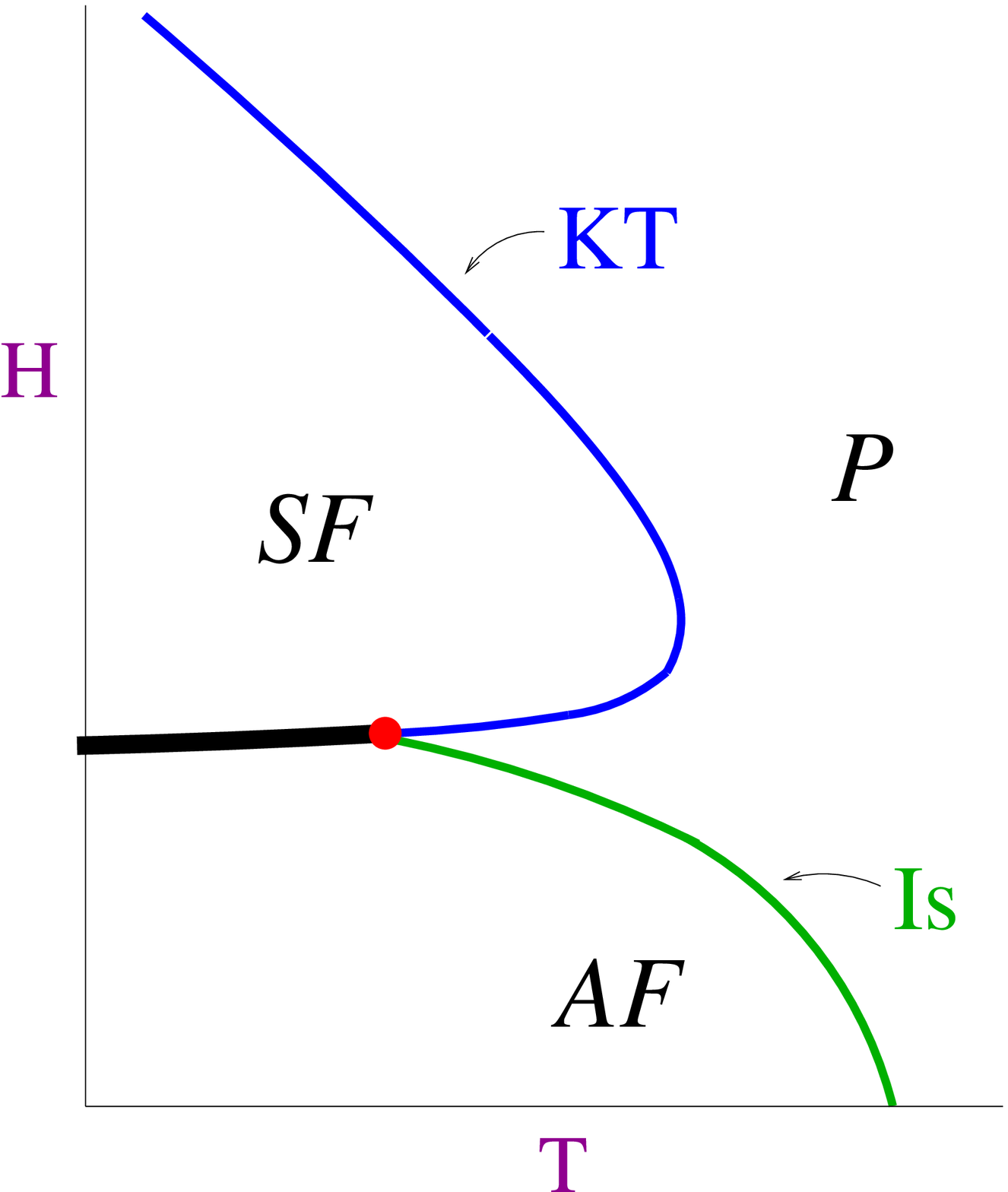}}
\end{minipage}
\hspace{1cm}
\begin{minipage}{7cm}
\centerline{\psfig{width=6truecm,angle=0,file=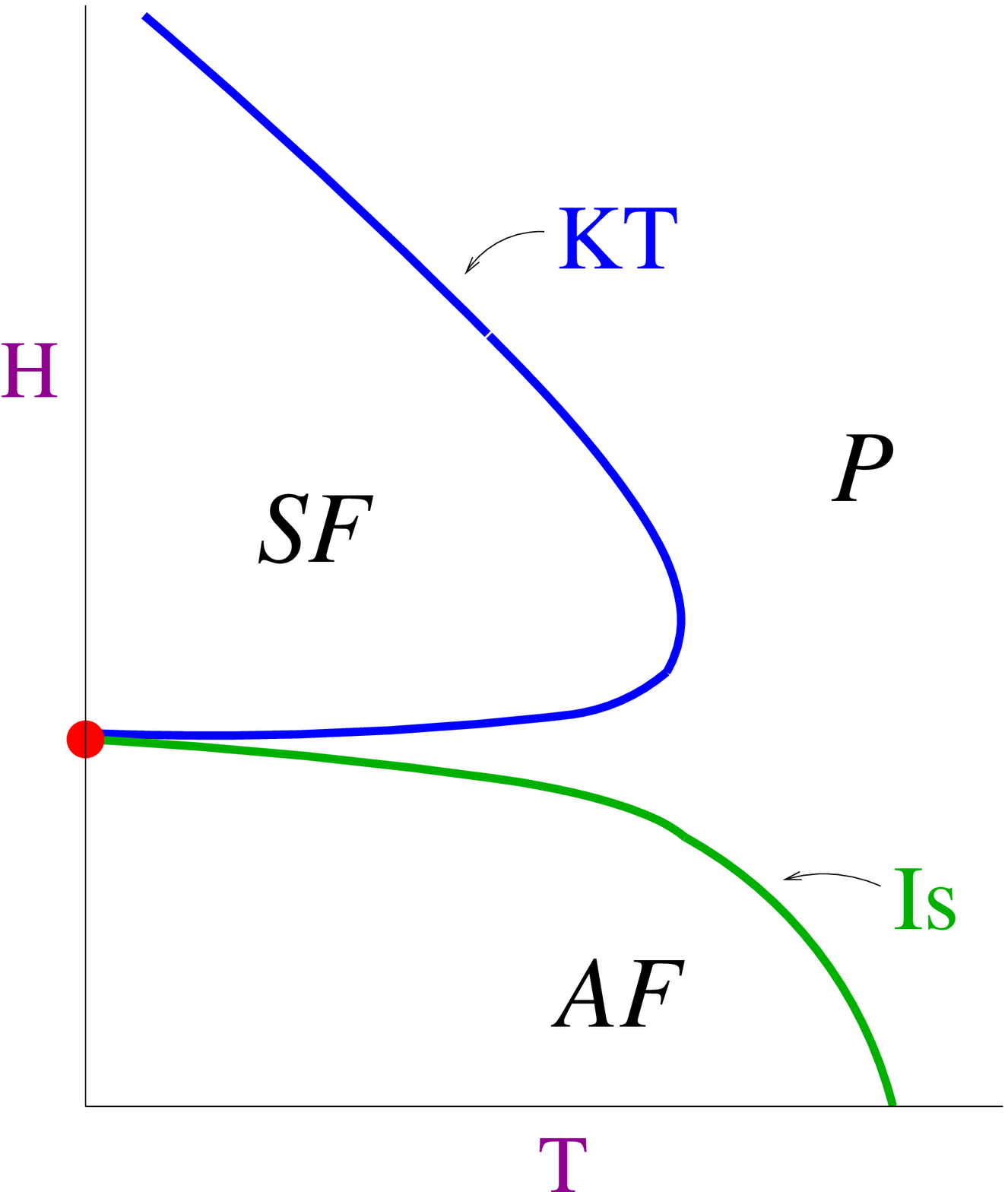}}
\end{minipage}
\vspace{2mm}
\caption{
  Phase diagrams in the $T$-$H$ plane in the presence of easy-axis
  anisotropy $A>0$ in 2D.
  The left figure shows a first-order transition line separating the
  antiferromagnetic (AF) and spin-flop (SF) phases, with a bicritical point
  where the Ising (Is) and the XY Kosterlitz-Thouless (KT) critical lines meet.  In
  the right figure the Ising and KT lines meet at a zero-temperature O(3)-symmetric
  MCP.  }
\label{ha}
\end{figure*}

The nature of the MCP is even more controversial in two dimensions
(2D) where different scenarios have been put forward; see, for
example, Refs.
\onlinecite{DRRH-82,DD-85,GJ-86,REPRKG-86,CABPST-93,KST-98,CLBE-01,LB-78,LB-81,CP-03,HSL-05,ZLS-06}.
The existence of the Ising and the XY Kosterlitz-Thouless (KT) critical lines
is well established. But, other features, like the existence of the 
spin-flop line, the nature of the MCP, and the MCP temperature, are still
the object of debate. For example, the two different phase diagrams
sketched in Fig.~\ref{ha} have apparently been both supported by
recent numerical Monte Carlo analyses: Ref.~\onlinecite{HSL-05} claims that MC data
are in agreement with the phase diagram reported on the left, while
Ref.~\onlinecite{ZLS-06} supports that reported on the right.  Similar
contradictions appear in the analysis of the experimental data.
\cite{DRRH-82,DD-85,GJ-86,REPRKG-86,CABPST-93,KST-98,CLBE-01}

In this paper we study the nature of the MCP in 2D within the field
theoretical approach.  Our main findings are the following:
\begin{itemize}
\item[(i)] The MCP cannot be O(3) symmetric and 
cannot occur at $T=0$.
Indeed, the magnetic field and the anisotropy give rise to an
{\em infinite} number of relevant perturbations of the O(3) FP. 
Moreover, since at $T=0$ the order parameter is discontinuous as $H$ 
is varied, we expect a first-order spin-flop line. We thus 
predict a finite-temperature MCP and 
exclude a phase diagram such as the one shown in Fig.~\ref{ha} on the right. 
\item[(ii)] We study the limit of weak anisotropy. In this limit we determine
how the critical temperature of the Ising and KT transition lines and of the MCP
vary with $H$ and $A$.
\end{itemize}

Our results on the phase diagram of the classical XXZ model (\ref{xxz}) are
also relevant for the phase diagram of quantum spin-$S$ XXZ models
\cite{CRTVV-01,CRTVV-03} and related systems, such as the hard-core boson
Hubbard model.\cite{HBSSTD-01,STTD-02}

We finally mention that anisotropic antiferromagnets modelled by
the XXZ model (\ref{xxz}) on a triangular lattice or on a stacked triangular 
lattice (and, more generally, on lattices that are not bipartite) are
expected to show a different multicritical behavior, essentially because
of frustration. Ref.~\onlinecite{CPV-05} reports a study of the possible phase
diagrams in the mean-field approximation and a field-theory study of the 
3D renormalization-group (RG) flow.

The paper is organized as follows. In Sec.~\ref{sec-iso} we summarize
some general predictions for the isotropic model. In Sec.~\ref{2dmc}
we investigate the stability of the O(3) FP by using field theory and
give predictions for the critical temperature as a function of $H$ and
$A$ for small $A$ and $H$. Finally, in Sec.~\ref{phdia} we discuss the
implications of the field-theoretical results and discuss the
possibile phase diagrams that are compatible with them.

\section{The isotropic antiferromagnet in two dimensions} \label{sec-iso}

For $H = A = 0$, the critical behavior of model (\ref{xxz}) is well known.
Indeed, if we perform the change of variables 
\begin{equation}
  \vec{\sigma}_n = (-1)^{[n]} \vec{S}_n,
\label{mapping-F-AF}
\end{equation}
where, in 2D, $[n] \equiv n_x + n_y$ [$n = (n_x,n_y)$], we obtain the 
ferromagnetic Heisenberg model for which several results are known.
\cite{PV-review} In particular, if we define the 
two-point function
\begin{equation}
  G(n) = \langle \vec{S}_0\cdot \vec{S}_n \rangle,
\end{equation}
a universal critical behavior is observed for the 
staggered susceptibility
\begin{equation}
\chi_s = \sum_n (-1)^{[n]} G(n)
\end{equation}
and for the staggered second-moment correlation length
\begin{equation}
\xi_s^2 = {1\over 4\chi_s} \sum_n (-1)^{[n]} |n|^2 G(n).
\end{equation}
Under the mapping (\ref{mapping-F-AF}), $\chi_s$ and $\xi_s$ 
go over to the standard 
susceptibility and correlation length of the ferromagnetic model. 
In 2D, perturbation theory together with the RG allows one
to derive the critical behavior of these two quantities for 
$T \to 0$.\cite{ZJ-book}
Using the results of Refs.~\onlinecite{CP-95,ACPP-99}, we obtain
\begin{eqnarray}
\chi_s &=& C_\chi \xi_s^2 \left({T\over 2\pi}\right)^2 
    \left[ 1 + 0.18169 T + 0.1334 T^2 + 0.1346 T^3 + O(T^4)\right],
\nonumber \\
\xi_s &=& C_\xi e^{2\pi/T}  \left({T\over 2\pi}\right)
  \left[ 1 - 0.0914 T - 0.1969 T^2 + O(T^3)\right].
\label{as-chi-xi}
\end{eqnarray}
The constants $C_\xi$ and $C_\chi$ cannot be determined in perturbation theory.
Numerical values are reported in Ref.~\onlinecite{CPRV-97}:
\begin{equation}
C_\chi = 93.25(3), \qquad C_\xi =  0.0124783(12) .
\end{equation}
The asymptotic expansions (\ref{as-chi-xi}) are quite accurate for 
$T\lesssim 0.3$, within a few percent at most.\cite{CEPS-95,CEMPS-96}

\section{Critical and multicritical behavior for small $A$ and $H$ }
\label{2dmc}

\subsection{General results} 

We consider the O(3) isotropic model at $H = 0$ and add terms that break 
the O(3) symmetry down to ${\mathbb Z}_2\otimes{\rm O}(2)$ 
(for instance the magnetic field or the 
anisotropy). The corresponding general Hamiltonian is
\begin{equation}
{\cal H}_{\rm gen} = J \sum_{\langle nm\rangle} 
    \vec{S}_n\cdot \vec{S}_m + \lambda Q(S).
\label{Hamiltonian-1}
\end{equation} 
If $Q(S)$ is a relevant perturbation of the O(3) FP, the O(3) critical point
is a MCP in the full theory. In 3D we can write
the singular part of the free energy\cite{KNF-76} for $\lambda\to 0$ as 
\begin{equation}
{\cal F}_{\rm sing} \sim u_t^{2-\alpha} B(X), \qquad X=u_\lambda u_t^{-\phi},
\label{scalmc}
\end{equation}
where $\alpha$ and $\phi$ are the O(3) 
specific-heat and crossover exponents, respectively, 
$B(X)$ is a universal scaling function, and $u_t$ and $u_\lambda$ 
are the scaling 
fields associated with the temperature and with $\lambda$.
In general, we expect 
\begin{equation} 
u_t = t + k \lambda, 
\end{equation}
where $k$ is a constant, $t \equiv T/T_{O(3)} - 1$ is the reduced
temperature, and $T_{O(3)}$ is the critical temperature of the
isotropic model. No such mixing between $t$ and $\lambda$ occurs in
$u_\lambda$, since $u_\lambda$ vanishes for $\lambda = 0$. Hence, we
can take $u_\lambda = \lambda$.  The crossover exponent is related to
the RG dimension $y_\lambda$ of the operator $Q$ that represents the
perturbation of the MCP: $\phi \equiv y_\lambda \nu$. Suppose now that
the system has a critical transition for $\lambda\not=0$ at
$T_c(\lambda)$. Since the singular part of the free energy close to a
critical point behaves as $(T - T_c)^{2-\alpha}$, we must have $B(X_c)
= 0$, where $X_c$ is the value of $X$ obtained by setting $T =
T_c(\lambda)$.  This equation is solved by $X_c = X_\pm$, where
$X_\pm$ are two constants that depend on the sign of $\lambda$, such
that $X_+ > 0$, $X_- < 0$. Hence, we obtain
\begin{equation}
\lambda \left[ {T_c(\lambda)\over T_{O(3)}} - 1 + k \lambda\right]^{-\phi} = X_\pm.
\end{equation}
It follows
\begin{eqnarray}
T_c(\lambda) &=& T_{O(3)}[1  + (\lambda/X_+)^{1/\phi} - k \lambda + \cdots],
\qquad  \lambda > 0,
\nonumber \\
T_c(\lambda) &=& T_{O(3)}[1  + (\lambda/X_-)^{1/\phi} - k \lambda + \cdots],
\qquad \lambda < 0.
\label{T-beh-0}
\end{eqnarray}
These expressions provide the $\lambda$ dependence of the critical
temperature for $\lambda$ small.  Note that, depending on the sign of
$\lambda$, $T_c(\lambda)$ varies differently.  The sign of $\lambda$
may also be relevant for the nature of the phase transition.  Indeed,
the low-temperature phase may be different depending on this sign: in
this case one observes critical behaviors belonging to different
universality classes for $\lambda > 0$ and $\lambda < 0$.

One can generalize these considerations to the case in which there are two relevant
perturbations $Q_1$ and $Q_2$ with parameters $\lambda_1$ and $\lambda_2$. In this
case we can write 
\begin{equation}
{\cal F}_{\rm sing} \sim u_t^{2-\alpha} B(X_1,X_2), 
\label{scal_2X}
\end{equation}
with 
\begin{eqnarray}
&& X_1 \equiv u_1 u_t^{-\phi_1}, \quad X_2 \equiv u_2 u_t^{-\phi_2}, \nonumber \\
&& u_1 = \lambda_1 + c_1 \lambda_2, \quad u_2 = \lambda_1 + c_2 \lambda_2,\nonumber \\
&& u_t = t + k_1 \lambda_1 + k_2 \lambda_2, 
\end{eqnarray}
where $k_1$, $k_2$, $c_1$, and $c_2$ are constants.
We have assumed here that there are two RG relevant operators that break 
the O(3) invariance, with RG dimensions $y_1 > y_2 > 0$, as is the 
case for anisotropic systems in the presence of a magnetic field.
The crossover exponents are $\phi_1 = \nu y_1$ and
$\phi_2 = \nu y_2$. The scaling fields $u_1$ and $u_2$ are linear combinations 
(and, beyond linear order, generic functions) of $\lambda_1$ and $\lambda_2$,
due to the fact that the lattice operators $Q_1$ and $Q_2$ 
generically couple both RG operators.

Let us now assume that, for $\lambda_1$ and $\lambda_2$ small, the model shows 
generically two types of phase transitions belonging to 
different universality classes
and that, for specific values of the ratio $\lambda_1/\lambda_2$,  
multicritical transitions occur. 
As before, we wish to derive the dependence of the 
critical temperature as a function of $\lambda_1$ and $\lambda_2$ for 
$\lambda_1,\lambda_2\to 0$. Since
\begin{equation}
X_2 = u_2 |u_1|^{-\phi_2/\phi_1} |X_1|^{\phi_2/\phi_1},
\end{equation}
we can rewrite Eq.~(\ref{scal_2X}) as 
\begin{equation}
{\cal F}_{\rm sing} \sim u_t^{2-\alpha} {\cal B}_\pm(X_1,u_2 |u_1|^{-\phi_2/\phi_1}), 
\label{scal_2X2}
\end{equation}
where we have introduced two different functions depending on the sign of $u_1$. 
If $u_1 > 0$ the relevant function is ${\cal B}_+(x,y)$, if $u_1 < 0$ one should 
consider ${\cal B}_-(x,y)$. At the critical point we must have 
\begin{equation} 
   {\cal B}_\pm(X_{1c},u_2 |u_1|^{-\phi_2/\phi_1}) = 0,
\end{equation}
which implies 
\begin{equation} 
X_{1c} = F_\pm (u_2 |u_1|^{-\phi_2/\phi_1}),
\end{equation}
with $F_+(x) > 0$ and $F_-(x) < 0$.
We obtain finally
\begin{equation}
  {T_c\over T_{O(3)}} - 1 \approx 
\left[ {u_1 \over F_\pm ( u_2 |u_1|^{-\phi_2/\phi_1} )} \right]^{1/\phi_1} - 
       k_1 \lambda_1 - k_2 \lambda_2.
\label{Tbeh-4}
\end{equation}
Let us now discuss some limiting cases. First, assume that 
$u_2 |u_1|^{-\phi_2/\phi_1} \ll 1$. This implies that $X_2$ is small
compared to $X_1$, so that we can neglect $X_2$. We are back to the case considered
before and thus we can use Eq.~(\ref{T-beh-0}). Depending on the sign of $u_1$ we have:
\begin{itemize}
\item[1)] If $u_1 > 0$, the phase transition is located at 
\begin{equation}
  {T_c\over T_{O(3)}} - 1 \approx \left( {u_1\over X_+} \right)^{1/\phi_1} - 
       k_1 \lambda_1 - k_2 \lambda_2,
\label{Tbeh-1}
\end{equation}
where $X_+ > 0$. 
\item[2)] If $u_1 < 0$, there is a phase transition located at 
\begin{equation}
  {T_c\over T_{O(3)}} - 1  \approx  \left( {u_1\over X_-} \right)^{1/\phi_1} - 
       k_1 \lambda_1 - k_2 \lambda_2,
\label{Tbeh-2}
\end{equation}
where $X_- < 0$. 
\end{itemize}
Comparing with Eq.~(\ref{Tbeh-4}) we obtain $F_\pm(0) = X_\pm$.

The second interesting limiting case corresponds to $u_1 \to 0$. In this case 
$X_1$ can be neglected and we need to consider only $X_2$. Therefore, depending
on the sign of $u_2$, we have a MCP located at 
\begin{equation}
  {T_c\over T_{O(3)}} - 1 \approx \left( {u_2\over X_{\rm mc,\pm}} \right)^{1/\phi_2} - 
       k_1 \lambda_1 - k_2 \lambda_2,
\label{Tbeh-3}
\end{equation}
where $X_{\rm mc,+} > 0$ and $X_{\rm mc,+} < 0$. Consistency with Eq.~(\ref{Tbeh-4})
requires 
\begin{eqnarray} 
   F_\pm(x) \approx \pm (x/X_{\rm mc,+} )^{-\phi_1/\phi_2} && \quad 
   \hbox{for $x\to +\infty$} , \nonumber \\
   F_\pm(x) \approx \pm (x/X_{\rm mc,-} )^{-\phi_1/\phi_2} && \quad 
   \hbox{for $x\to -\infty$}.
\end{eqnarray}
One may devise simple interpolations that are exact for $x\to 0$ and $x\to\infty$.
If, for instance, $u_1 > 0$ and $u_2 > 0$,
we may consider the approximate expression
\begin{equation}
 {T_c\over T_{O(3)}} - 1 \approx
\left( {u_1 + b u_2^{\phi_1/\phi_2} \over X_+} \right)^{1/\phi_1} - 
       k_1 \lambda_1 - k_2 \lambda_2, 
\qquad b \equiv X_+ X_{\rm mc,+}^{-\phi_1/\phi_2}.
\label{Tbeh-5}
\end{equation}
The functions $F_\pm(x)$ are crossover functions that interpolate between the two 
regimes in which only one of the relevant operators is present. 
As far as the nature of the transition, the relevant quantity is the sign of $u_1$. 
In general, we expect that the transition belongs to different universality classes
depending on the sign of $u_1$. For $u_1 = 0$ the leading relevant operator decouples
and thus we obtain a MCP whose nature may depend on the sign of $u_2$.

The previous results apply to the 3D model but cannot be used directly 
in 2D, since in this case
$T_{O(3)} = 0$ and $\nu$ is not defined
($\xi_s\sim e^{2\pi/T}$ for $\lambda=0$). To investigate the 2D case,
let us consider again Hamiltonian (\ref{Hamiltonian-1}), let us assume that 
$Q(S)$ renormalizes multiplicatively under RG transformations and that its
RG dimension is 2 with logarithmic corrections 
(as we shall see, this is the case of interest). 
The perturbative analysis of the scaling behavior of the 
free energy is analogous to that presented in Ref.~\onlinecite{CMP-01}. The 
correct scaling variable is 
\begin{equation}
Y = u_\lambda \left({T\over2\pi}\right)^p e^{4\pi/T} \sim 
   u_\lambda \left({T\over2\pi}\right)^{p-2} \xi_s^2,
\end{equation}
where we used Eq.~(\ref{as-chi-xi}), and 
$p$ is a power that can be computed by using the one-loop 
expression of the anomalous dimension of $Q$. Then, the singular part of the 
free energy is given by
\begin{equation}
{\cal F}_{\rm sing} \approx \xi_s^{-2} {B}(Y).
\end{equation}
The critical line is again characterized by $Y = Y_\pm$, i.e. by
\begin{equation}
\lambda \left({T_c\over2\pi}\right)^p e^{4\pi/T_c} = Y_\pm,
\end{equation}
with $Y_+ > 0$ and $Y_- < 0$.
Solving this equation for $T_c$, we obtain
\begin{equation}
T_c = {4\pi\over \ln (Y_\pm/\lambda)} 
    \left[1 - p {\ln {1\over2}\ln (Y_\pm/\lambda) \over \ln (Y_\pm/\lambda)}
    + \cdots \right].
\end{equation}
The discussion is analogous in the case there are two relevant perturbations.
We define two scaling variables $Y_1$ and $Y_2$ as
\begin{eqnarray}
Y_i = u_i \left({T\over2\pi}\right)^{p_i} e^{4\pi/T},\quad i=1,2,
\end{eqnarray}
and write the free energy as 
\begin{equation}
{\cal F}_{\rm sing} = \xi_s^{-2} B(Y_1,Y_2).
\label{Fsing-2D}
\end{equation}
If $p_1 < p_2$, the critical behavior depends on the sign of $u_1$:
For $u_1 > 0$ and $u_1 < 0$ one obtains different critical behaviors.
In the limit in which $Y_2$ is small and can be neglected, we can write
\begin{equation}
T_c = {4\pi\over \ln (Y_\pm/u_1)} 
    \left[1 - p_1 {\ln {1\over2}\ln (Y_\pm/u_1) \over \ln (Y_\pm/u_1)}
    \right],
\label{T-beh2d-1}
\end{equation}
where $Y_\pm$ are two constants such that $Y_+ > 0$ and $Y_- < 0$.
As observed in the 3D case, 
Eq.~(\ref{T-beh2d-1}) [it is the analogue of Eqs.~(\ref{Tbeh-1}) and (\ref{Tbeh-2})]
holds only if $Y_2$ can be neglected. Since
\begin{equation} 
Y_2 = {u_2\over u_1} Y_1 \left({T\over2\pi}\right)^{p_2 - p_1} \approx
    {u_2\over u_1} Y_1 
    \left( {1\over 2} \ln (Y_1/u_1) \right)^{p_1 - p_2},
\end{equation}
Eq.~(\ref{T-beh2d-1}) holds only in the parameter region in which 
\begin{equation}
   u_2 \ll u_1 \left( {1\over 2} \ln (Y_\pm/u_1) \right)^{p_2 - p_1},
\label{cond-2d}
\end{equation}
i.e. far from the MCP at $u_1 = 0$.  For $|u_1|\to 0$ this
region shrinks to zero and thus a correct formula requires the full crossover
function.

For $u_1 = 0$, we have a MCP at 
\begin{equation}
T_c = {4\pi\over \ln (Y_{\rm mc,\pm}/u_2)} 
 \left[1 - p_2 {\ln {1\over2}\ln (Y_{\rm mc,\pm}/u_2) \over \ln (Y_{\rm mc,\pm}/u_2)}
    \right],
\label{T-beh2d-2}
\end{equation}
depending on the sign of $u_2$. 

\subsection{Effective Hamiltonians} \label{effective-Hamiltonians}

Let us now apply the results of the previous section to the 
XXZ model. For this purpose 
we consider the ferromagnetic Hamiltonian corresponding to
(\ref{xxz}) under the mapping (\ref{mapping-F-AF}),
\begin{equation}
{\cal H}_f = - \sum_{\langle nm\rangle}\vec{\sigma}_n\cdot \vec{\sigma}_m 
    -A \sum_{\langle nm\rangle} \sigma_{n,z} \sigma_{m,z} + 
    H \sum_n (-1)^{[n]} \sigma_{n,z},
\label{xxz-ferro}
\end{equation}
where we have set $J = 1$ for convenience.
We argue that Hamiltonian (\ref{xxz-ferro}) is equivalent to the Hamiltonian
\begin{equation}
{\cal H}^* = - \sum_{\langle nm\rangle} \vec{\sigma}_n\cdot \vec{\sigma}_m + 
    \sum_{l\ge 1} \alpha_l(T,h,a) {\cal O}_{2l}(\sigma),
\label{Hstar}
\end{equation}
where $ {\cal O}_{2l}(\sigma) $ are the zero-momentum dimension-zero 
spin-$2l$ perturbations of the O(3) FP, $h = H/T$, $a = A/T$ (note that 
the partition function depends on $h$ and $a$). The operators 
$ {\cal O}_{2l}(\sigma) $ can be constructed starting from the 
symmetric traceless operators of degree $2l$, see, e.g., Ref.~\onlinecite{CP-94}. 
Explicitly, for $l=1,2$ we have
\begin{eqnarray}
&& {\cal O}_{2}(\sigma) = \sum_n (\sigma_{n,z}^2 - {1/3}), \nonumber \\
&& {\cal O}_{4}(\sigma) = 
   \sum_n \left(\sigma_{n,z}^4 - {6\over7} \sigma_{n,z}^2 + 
      {3\over 35} \right).
\end{eqnarray}
The functions $\alpha_{l}(T,h,a)$ are smooth and vanish for $h\to 0$ and $a \to 0$.
It should be stressed that the two Hamiltonians are equivalent 
only for the computation of the leading critical behavior 
of long-distance quantities.  

The equivalence of ${\cal H}_f$ and ${\cal H}^*$ for what concerns the
critical behavior can be justified on the basis of the results of
Ref.~\onlinecite{KNF-76} obtained in the usual LGW approach.  In the absence
of anisotropy and magnetic field, i.e. for $A=H=0$, model (\ref{xxz}) is O(3)
invariant and thus its critical behavior is described by the usual LGW
Hamiltonian
\begin{equation}
{\cal H}_{\Phi^4} = \int d^dx\, \left[
     {1\over2} \sum_\mu (\partial_\mu \Phi)^2  +
     {r\over2} \Phi^2 + {u\over 4!} \Phi^4 \right],
\end{equation}
where $\vec{\Phi}$ is a three-component vector.
The magnetic field and the anisotropy break the O(3) invariance. If we define
$\vec{\Phi} = (\vec{\phi},\varphi)$,
where $\vec{\phi}$ is a two-component vector and 
$\varphi$ a scalar,
the LGW Hamiltonian (\ref{mcphi4}) 
corresponding to (\ref{xxz}) 
can be written as \cite{KNF-76}
\begin{eqnarray}
{\cal H} = H_{\Phi^4} + \int d^d x \Bigl[
 f_{22} P_{22} + f_{42} P_{42} + f_{44} P_{44}) \Bigr],
\label{mcphi4b} 
\end{eqnarray}
where 
\begin{eqnarray}
&&P_{22}= \varphi^2-{1\over 3} \Phi^2 ,\nonumber \\ 
&&P_{42}= \Phi^2 P_{22},\nonumber \\
&&P_{44}=  
\varphi^4 - {6\over 7}\Phi^2 \varphi^2 + {3\over 35} (\Phi^2)^2,
\end{eqnarray}
and $f_{ij}$ are coupling constants depending on $a$, $h^2$, and $T$.
We have introduced here the homogeneous polynomial $P_{ml}$. The polynomial 
$P_{ll}$ has degree $l$ and transforms irreducibly as a spin-$l$ 
representation of the O(3) group.\cite{footnote} 
Polynomials $P_{ml}$, $m>l$, are defined as 
$P_{ml} \equiv (\Phi^2)^{(m-l)/2} P_{ll}$.
The classification of the zero-momentum perturbations in terms of spin values 
is particularly convenient,
since polynomials with different spin do not mix under RG transformations and
the RG dimensions $y_{ml}$ do not depend on the particular component of the
spin-$l$ representation.  We refer to Ref.~\onlinecite{CPV-03} for details.
Thus, in the LGW approach the O(3) Hamiltonian is perturbed by spin-2 and spin-4
perturbations. If one considers higher power of the fields, also 
spin-6, spin-8, $\ldots$ perturbations appear. They are irrelevant in 3D, 
but must be considered in 2D, since in this case the field $\Phi$ is dimensionless
and any polynomial of the fields is relevant.

\subsection{Three-dimensional behavior}

In 3D the critical behavior is well known:
\begin{itemize}
\item[(i)] In the absence of anisotropy, i.e. for $A=0$,
there is an XY critical line $T_{XY}(H)$ 
ending at the O(3) critical point for $H=0$. 
The phase diagram is symmetric under $H\to -H$. 
\item[(ii)] In the absence of magnetic field, i.e. for $H=0$,
there are two critical lines: 
an XY critical line for $A < 0$ and an Ising critical line for $A > 0$,
which meet at the O(3) critical point as $A\to 0$.
\item[(iii)] If $A$ and $H$ are both present, one observes two different phase diagrams
depending on the sign of $A$. If $A$ is negative, there is 
an XY transition. If $A$ is positive, there is an Ising critical line 
for $H$ small and an XY critical line for $H$ large; the two lines meet 
at a MCP corresponding to a nonvanishing value of $H$. 
\end{itemize}
We wish now to determine the dependence of the critical temperatures on
$A$ and $H$ close to the O(3) point. For this purpose we investigate the 
relevance of the perturbations. In 3D we expect that only polynomials 
of degree two or four are relevant. Thus, $P_{22}$, $P_{42}$ and $P_{44}$
are the only quantities that should be considered. The RG analysis is 
reported in Ref.~\onlinecite{CPV-03} and indicates that $P_{42}$ is irrelevant, while 
$P_{22}$ and $P_{44}$ are relevant. Perturbative field-theory calculations
provide estimates of the corresponding RG dimensions:\cite{CPV-03,CPV-00}
$y_{22} \approx  1.79$, $y_{44} \approx  0.01$, and $y_{42}\approx -0.55$.
The relevant scaling fields are
$u_1 = a + c_1 h^2$ associated with $P_{22}$ and $u_2 = a + c_2 h^2$ associated with 
$P_{44}$. The scaling fields depend on $h^2$, because of the symmetry
under $H\to -H$ transformations.
Since $P_{22}$ is the most relevant operator, 
the critical behavior depends on the sign of $u_1$. We obtain 
XY behavior for $a + c_1 h^2 < 0$ and Ising behavior for $a + c_1 h^2  > 0$.
Since for $A= 0$ only the XY transition is observed,
the constant $c_1$ must be negative. Hence, Ising behavior can only be observed
for $a$ positive and $h^2$ not too large, in agreement with experiments.
The critical temperatures behave as 
\begin{eqnarray}
T_{c,\rm XY} - T_{O(3)} &=&  b_{\rm XY}  (-a - c_1 h^2)^{1/\phi_{22}} - 
         k_1 a - k_2 h^2, 
\nonumber \\
T_{c,\rm Is}- T_{O(3)} &=& b_{\rm Is}  (a + c_1 h^2)^{1/\phi_{22}} - 
         k_1 a - k_2 h^2,
\end{eqnarray}
where $b_{\rm XY}$, $b_{\rm Is}$, $k_1$ and $k_2$ are constants, with
$b_{\rm XY} > 0$, $b_{\rm Is} > 0$. Here \cite{CPV-03} $\phi_{22} = 1.260(11)$,
$1/\phi_{22} = 0.794(7)$.

Multicritical behavior is observed for $a = - c_1h^2$ (since $c_1$ is 
negative this equality can only occur for $a > 0$). Since 
$1/\phi_{44} \gtrsim 100$, the leading nonanalytic term with exponent $1/\phi_{44}$ 
cannot be observed in practice, and therefore
\begin{equation}
T_{c,\rm mc} - T_{O(3)} = - k_1 a - k_2 h^2.
\end{equation}
 
\begin{figure*}[tb]
\centerline{\psfig{width=8truecm,angle=0,file=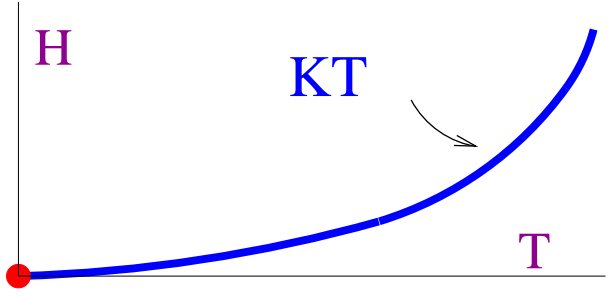}}
\vspace{2mm}
\caption{
Phase diagram of isotropic antiferromagnets in the $T$-$H$ plane. 
}
\label{noani}
\end{figure*}

\begin{figure*}[tb]
\centerline{\psfig{width=6truecm,angle=0,file=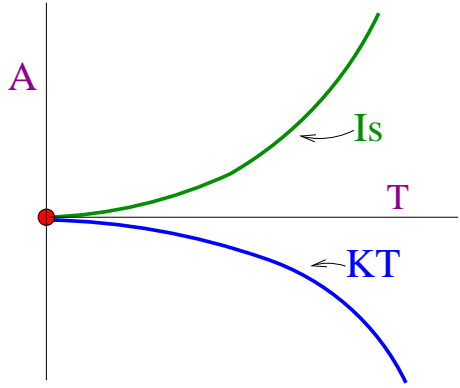}}
\vspace{2mm}
\caption{
Phase diagram of anisotropic antiferromagnets in the $T$-$A$ plane for $H=0$.
}
\label{noH}
\end{figure*}

\subsection{Two-dimensional behavior} \label{sec-2d-small}

The critical behavior in 2D is analogous to that observed in 3D.
For $A < 0$ there is an XY Kosterlitz-Thouless (KT) transition for any $H$,
while for $A > 0$ there is an Ising transition for small $|H|$, an XY
transition for large $|H|$, and a MCP in between.  The phase diagram in the
two limiting cases $A = 0$ and $H = 0$ is reported in Figs.~\ref{noani} and
\ref{noH}.

To compute the position of the critical point, we should investigate the
relevance of the perturbations $P_{ml}$.  As shown in Ref.~\onlinecite{BZL-76}, in
2D any spin-$l$ perturbation $P_{ll}$ is relevant at the O(3) FP, since the
corresponding RG dimension $y_{ll}$ is positive, indeed 
$y_{ll}=2$ apart from logarithms
which can be computed by using perturbation theory. One can also argue that
perturbations $P_{ml}$ with $m > l$ can be neglected. Indeed, spin waves,
that are rotations of the spins, are the critical modes of the ferromagnetic
O(3) model.  Changes in the size of the field $\Phi$ should not be critical.
Therefore, $P_{ml} \equiv (\Phi^2)^{(m-l)/2} P_{ll} $ should be equivalent
to $(\Phi_0^2)^{(m-l)/2} P_{ll} \sim P_{ll}$, where $\Phi_0^2$ is the
average of $\Phi^2$. Thus, one should only consider the operators $P_{ll}$.
Equation (\ref{Hstar}) then follows immediately.  The equivalence of model
(\ref{xxz}) with Hamiltonian (\ref{Hstar}) was already conjectured in
Ref.~\onlinecite{NP-77}, even though there only the leading spin-2 term was
explicitly considered.

The relevant scaling variables are
\begin{equation}
Y_{2l} = u_{l} \left({T\over 2\pi}\right)^{l (2l +1) + 2} e^{4\pi/T}, 
\label{scalingY}
\end{equation}
which are associated with the spin-$2l$ perturbation. 
The power of $T$, which is universal, has been determined
by using the perturbative results of Ref.~\onlinecite{BZL-76}, which provide
the anomalous dimension of any zero-dimension spin-$l$ perturbation of 
the 2D $N$-vector model. The scaling fields $u_{l}$ are 
linear combinations of the 
parameters that break the O(3) invariance. As before, we write them as
$u_{l} = a + c_{l} h^2$.
As a consequence, the free energy can be written as 
\begin{equation}
{\cal F}_{\rm sing} \approx \xi^{-2} 
     \hat{F} (Y_2,Y_4,\ldots).
\end{equation}
Note that different powers of $T$ appear in the definition (\ref{scalingY}). 
The most relevant term for $T\to 0$ corresponds to the spin-2 operator, since 
$Y_{2l} \sim Y_2 T^{l (2l + 1) - 3}$ and $l (2l + 1) - 3$ is positive 
for $l \ge 2$. Next one should consider the spin-4 scaling variable $Y_4$.
If one considers the scaling limit at fixed $Y_2\not = 0$ or 
$Y_4\not = 0$, 
the higher-order spin variables go to zero as $T\to 0$ and thus represent 
corrections to scaling proportional to powers of $T$. 
Thus, in the scaling limit one can neglect $Y_6$, $Y_8$, $\ldots$, and write
\begin{equation}
{\cal F}_{\rm sing} \approx \xi^{-2} 
     F_2 (Y_2,Y_4),
\label{FsingY2}
\end{equation}
which coincides with Eq.~(\ref{Fsing-2D}). We can thus use the above-obtained 
results. 
\begin{itemize}
\item[(i)] For $u_1 < 0$, i.e. $a + c_1 h^2 < 0$, there is a KT transition at
\begin{equation}
T_{c,\rm XY} = {4\pi\over \ln (Y_{\rm XY}/u_1)} 
    \left[1 - 5 {\ln {1\over2}\ln (Y_{\rm XY}/u_1) \over \ln (Y_{\rm XY}/u_1)}
    \right],
\label{T2d-1}
\end{equation}
where $Y_{\rm XY} < 0$. Note that the occurence of an XY transition for 
$a = 0$ implies $c_1 < 0$ as in the 3D case.
\item[(ii)] For $u_1 > 0$, i.e. $a + c_1 h^2 > 0$, there is an Ising transition at
\begin{equation}
T_{c,\rm Is} = {4\pi\over \ln (Y_{\rm Is}/u_1)} 
    \left[1 - 5 {\ln {1\over2}\ln (Y_{\rm Is}/u_1) \over \ln (Y_{\rm Is}/u_1)}
    \right],
\label{T2d-2}
\end{equation}
where $Y_{\rm Is} > 0$. Since $c_1 < 0$  this transition can only occur for $a > 0 $
and $h^2$ small.
\item[(iii)] For $u_1 = 0$, i.e. $a + c_1 h^2 = 0$, there is a MCP at
\begin{equation}
T_{c,\rm mc} = {4\pi\over \ln (Y_{\rm mc}/h^2)} 
    \left[1 - 12 {\ln {1\over2}\ln (Y_{\rm mc}/h^2) \over \ln (Y_{\rm mc}/h^2)}
    \right],
\end{equation}
where $Y_{\rm mc} > 0$. Note that since the MCP occurs
for $a = - c_1 h^2$, we can replace $u_2 = a + c_2 h^2$ simply with $h^2$.
\end{itemize}
Note that, as discussed before, Eqs.~(\ref{T2d-1}) and (\ref{T2d-2})
are valid only as long as condition (\ref{cond-2d}) is satisfied, i.e.
far from the MCP $u_1 = 0$.

\section{Discussion and conclusions}
\label{phdia}

\begin{figure*}[tb]
\centerline{\psfig{width=9truecm,angle=0,file=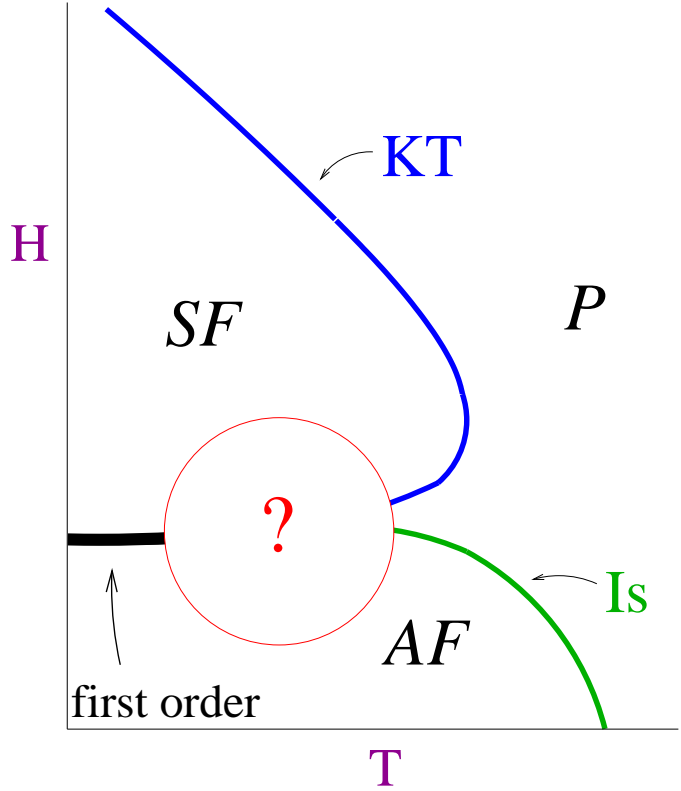}}
\vspace{2mm}
\caption{
Phase diagram in the $T$-$H$ plane in the presence of uniaxial 
anisotropy $A>0$.
}
\label{hau}
\end{figure*}

\begin{figure*}[tb]
\centerline{\psfig{width=10truecm,angle=0,file=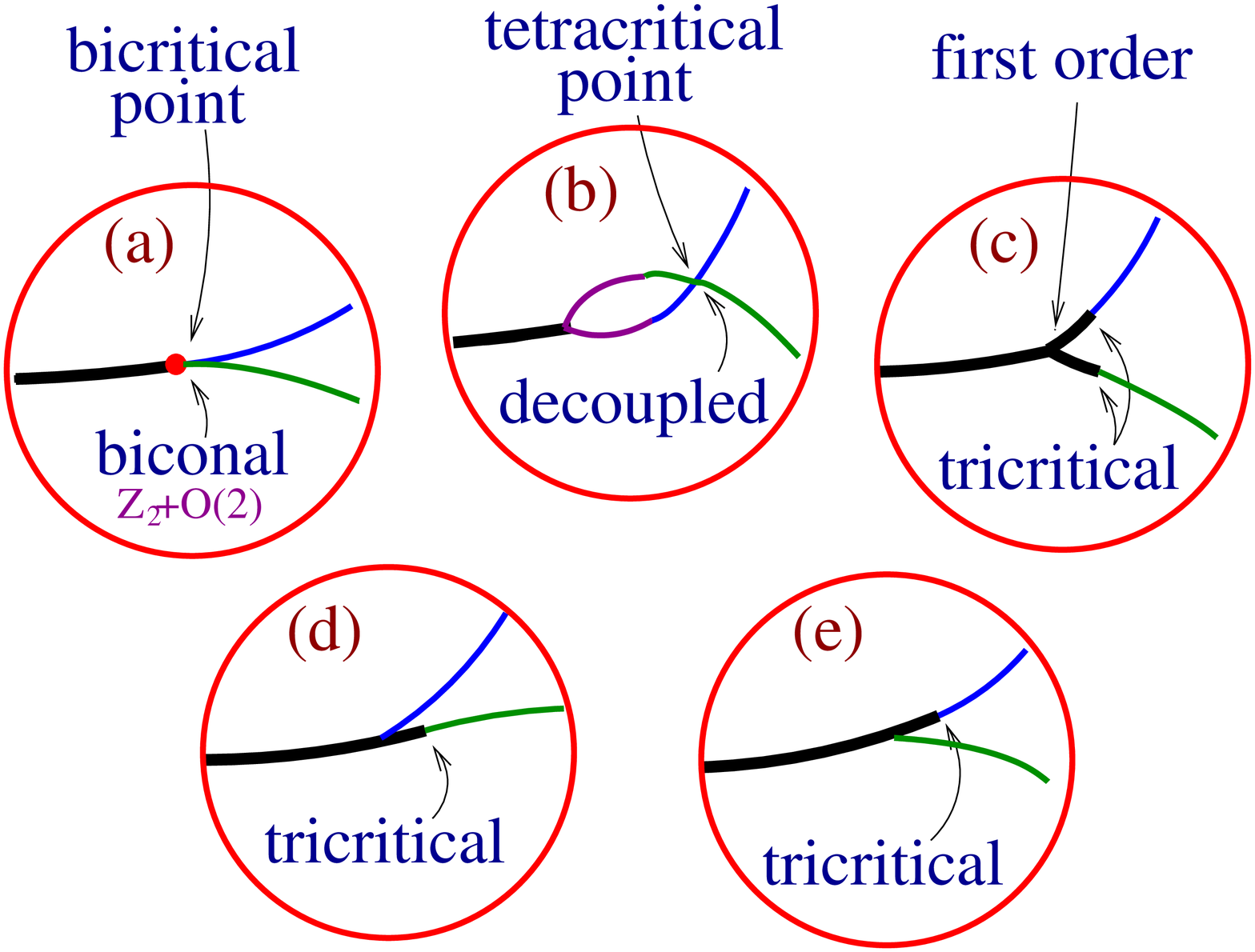}}
\vspace{2mm}
\caption{
Some possible phase diagrams inside the blob
appearing in Fig.~\ref{hau}.
}
\label{haublob}
\end{figure*}

In Sec.~\ref{sec-2d-small} we have discussed the behavior of the
classical 2D XXZ model (\ref{xxz}) on a square lattice in the presence of a
magnetic field along the easy axis, for $A$ and $H$ small, focusing on
the dependence of the critical temperature on $A$ and $H$. In this section we wish to
discuss the possible scenarios for the nature of the MCP.

First, we exclude the possibility that the MCP is O(3)
symmetric and located at $T = 0$, and thus we exclude a phase diagram
such as the one shown on the right in
Fig.~\ref{ha}. This is implicit in the results of 
Sec.~\ref{sec-2d-small}, but we wish to present here an extended discussion.
There are essentially two observations that exclude an O(3) MCP.
\begin{itemize}
\item  For $T=0$ 
the XXZ model (\ref{xxz}) on a square lattice can be easily solved.
\cite{BRW-71,LB-78}  At fixed $A>0$, one finds two critical values of the 
magnetic field: 
\begin{eqnarray}
&&H_{c1} = 4 J \sqrt{ 2 A + A^2},\nonumber\\
&&H_{c2} = 4 J (2 + A).
\label{hc}
\end{eqnarray}
For $|H|<H_{c1}$ 
the system is in a fully aligned antiferromagnetic configuration, for 
$H_{c1} < |H| <H_{c2}$ the system is in a
spin-flop configuration, while for $|H|>H_{c2}$  all spins are aligned with 
the magnetic field.  The spin-flop transition at $|H| = H_{c1}$ is of first order,
since the order parameter, the staggered magnetization, has a discontinuity.
If the transition were at $T=0$, the MCP should coincide
with the spin-flop point $|H|=H_{c1}$. Thus, the magnetization, the susceptibility
and all critical quantities would have a discontinuity at $T=0$ when varying $H$.
It is unclear how an O(3) critical behavior might be consistent with 
this discontinuity. On the other hand, the first-order behavior at $T=0$
is consistent with the existence of a first-order spin-flop line.
\item The second objection is based on the LGW analysis presented in 
Sec.~\ref{effective-Hamiltonians}. The O(3) FP is unstable under an 
{\em infinite} number of perturbations. Thus, an infinite number of tunings 
is needed to recover the O(3) symmetry.
\end{itemize}
On the basis of these remarks, in 2D anisotropic antiferromagnets in a
magnetic field along the easy axis, a scenario based on a $T=0$ O(3)-symmetric
MCP appears untenable. 
These conclusions are analogous to those that hold in 3D.\cite{CPV-03,HPV-05}
It should be noted that in 2D the argument is much stronger. While in 3D
these conclusions are based on the numerical determination of the RG dimension
of the spin-4 perturbation, in 2D they follow from the relevance of the spin-$l$ 
perturbations, which is an {\em exact} result. Moreover, while in 3D 
the O(3) behavior can be obtained by tuning a single additional parameter
in such a way to decouple both the spin-2 and the spin-4 operator, in 2D 
the O(3) behavior can never be obtained at finite $A$, since an 
infinite number of tunings is needed.

Since we have proved that in 2D the MCP cannot have O(3) symmetry,
an open question concerns the nature of the MCP. We shall now 
show that the decoupled FP, corresponding to a
multicritical behavior in which the two order parameters are effectively uncoupled, 
is stable.
The stability of the decoupled FP can be proved by nonperturbative
arguments.\cite{Aharony-02} Indeed, the RG dimension $y_w$ of the operator
$\Phi_1^2 \Phi_2^2$ that couples the two order parameters in the LGW theory (\ref{mcphi4})
is
given by
\begin{equation}
y_w =  {1\over \nu_{\rm KT}} + {1 \over \nu_{\rm Is}} - 2=-1<0.
\label{yw}
\end{equation}
since $\nu_{\rm KT}=\infty$ and $\nu_{\rm Is}=1$.  Therefore, the
perturbation is irrelevant and the decoupled FP is stable.  Note that
a decoupled MCP is always tetracritical, as in
Fig.~\ref{tetra}. 

The stability of the decoupled FP and the instability of the O(3) FP
is also consistent with some general arguments.\cite{VZ-06} At a MCP
the exponent $\eta$ describing the critical behavior of the
correlation function of the order parameter is replaced by a matrix
$\eta_{ij}$, see, e.g., Ref.~\onlinecite{KNF-76}.  The conjecture of
Ref.~\onlinecite{VZ-06} states that ${\rm Tr}\,\eta$ should have a
maximum at the stable FP. This indeed occurs in the present case:
at the decoupled FP we have ${\rm Tr}\,\eta = 3/4$ while at the
O(3) FP ${\rm Tr}\,\eta = 0$.

In Fig.~\ref{hau} we show a plausible phase diagram, with three
transition lines: a spin-flop first-order transition line, an 
Ising and a KT critical line.  The phase diagram inside the blob 
is an open issue.  
Some possibilities are shown in Fig.~\ref{haublob}:

\begin{itemize}

\item
Fig.~\ref{haublob} (a) presents a bicritical point, which may be associated
with a stable biconal FP, whose 
attraction domain is in the bicritical region of the bare parameters
of the LGW $\Phi^4$ theory. We should say that we do not have any
evidence for the existence of such a FP.

\item
In Fig.~\ref{haublob} (b) we show a tetracritical point, where the Ising
and KT lines intersect each other. In this case the 
MCP may be controlled by the stable decoupled FP discussed above.

\item
In Fig.~\ref{haublob} (c) the transition at the MCP is of first order.
Starting from the MCP, the first-order transitions extend up to 
tricritical points, where the Ising and the KT critical lines start.
This occurs if the system is outside the attraction domain of the
stable FP of the RG flow. This scenario resembles the one predicted 
in 3D, see Fig.~\ref{tricri}.

\item Finally, we cannot exclude phase diagrams like those shown in 
  Fig.~\ref{haublob} (d) and Fig.~\ref{haublob} (e).
  Case (d) is apparently observed  in 
  antiferromagnets with single-ion anisotropy and more than nearest-neighbor
  interactions,\cite{LS-04} and also in hard-core boson systems,\cite{STTD-02} 
   which are
  equivalent to anisotropic spin-1/2 XXZ systems in a magnetic field.

\end{itemize}

Of course, further experimental and theoretical
investigations are called for to conclusively settle
this issue.

\acknowledgments
We thank Pasquale Calabrese for useful discussions.


\begin{thebibliography}{99}
\bibitem{footnote}
Explicitly, we can take $P_{ll} \sim  (\Phi^2)^{l/2} P_l(x)$,
where $x^2 = \varphi^2/\Phi^2$ and $P_l(x)$ is a Legendre 
polynomial.
\end{thebibliography}
\end{document}